\documentclass[aps,prl,twocolumn]{revtex4}
\usepackage{graphicx}
\usepackage{latexsym}
\usepackage{amsmath}
\usepackage{amsfonts}
\usepackage{amssymb}
\newcommand{\be}{\begin{equation}}
\newcommand{\ee}{\end{equation}}
\newcommand{\bea}{\begin{eqnarray}}
\newcommand{\eea}{\end{eqnarray}}
\begin{document}
\title{Worm-like Polymer Loops and Fourier Knots}
\author{S.M. Rapaport$^{\dag}$, Y. Rabin$^{\dag}$ and A. Yu. Grosberg$^{\ddag}$}
\affiliation{$^{\dag}$Department of Physics, Bar--Ilan University, Ramat-Gan 52900, Israel }
\affiliation{$^{\ddag}$Department of Physics, University of Minnesota, 116 Church Street
SE, Minneapolis, MN 55455, USA}

\begin{abstract}
Every smooth closed curve can be represented by a suitable Fourier
sum. We show that the ensemble of curves generated by randomly
chosen Fourier coefficients with amplitudes inversely proportional
to spatial frequency (with a smooth exponential cutoff), can be
accurately mapped on the physical ensemble of worm-like polymer
loops. We find that measures of correlation on the scale of the
entire loop yield a larger persistence length than that calculated
from the tangent-tangent correlation function at small length
scales. The conjecture that physical loops exhibit additional
rigidity on scales comparable to the entire loop due to the
contribution of twist rigidity, can be tested experimentally by
determining the persistence length from the local curvature and
comparing it with that obtained by measuring the radius of
gyration of dsDNA plasmids. The topological properties of the
ensemble randomly generated worm-like loops are shown to be
similar to that of other polymer models.

\end{abstract}
\maketitle

Classical polymer theories, from Debye to Flory to De Gennes
\cite{DeGennes} and beyond, make extensive use of the analogy
between conformation of a chain molecule and a Brownian random
walk (BRW). In mathematically idealized form, such a random walk
is thought of as a Wiener trajectory $\vec{r}(\tau)$ generated by
the measure $P\left\{  \vec{r}(\tau)\right\}  \propto\exp\left[
-\mathrm{const} {\displaystyle\int} \left(
\partial\vec{r}/\partial\tau\right)  ^{2}d\tau\right]$, where
$\tau$ is a parameter running along the trajectory. The BRW model
brings powerful mathematical techniques, such as the diffusion
equation, to bear on polymers but, however fruitful, this method
has its limitations. For example, it is fundamentally unable to
reproduce the fact of finite extensibility of polymer chains
stretched by a strong force\cite{MarcoSiggiaBustamante}. Another
field in which the Wiener trajectory model fails miserably is the
study of polymers with knots. Indeed, simulations of discrete
polymer models, or random polygons with $N$ steps, show
\cite{N01,N02,N03}, that the probability of trivial knot
configuration decays exponentially with $N$ as
$P_{0}(N)\sim\exp\left( -N/N^{\ast}\right)  $, where $N^{\ast}$
defines the crossover from an unknotted to a knotted regime. It
was argued that Wiener trajectory polymer models can not be used
to calculate this probability \cite{grosberg}, since they
correspond to the limit $N\rightarrow\infty$, $l\rightarrow0$ ($l$
is a step length); $Nl^{2}=\mathrm{const}$, in which the
probability of a trivial knot vanishes. Notice that the contour
length of Wiener trajectory diverges, $Nl\rightarrow\infty$, so it
is not surprising that these models do not exhibit finite
extensibility. A better continuum model of polymers which can
handle the constraints imposed both by finite extensibility and by
the presence of knots, and can describe the elastic response of
such objects, is that of a worm-like chain. However, while the
statistical physics of linear worm-like polymers is well
understood, little is known about the conformations and the
topology of worm-like loops (WLL). Our goal in this article is to
work out a method allowing computational generation of an ensemble
of smoothly curved conformations of such polymers and to carry out
the statistical analysis of their geometric and topological
properties.

The conformations of worm-like polymers are generated by the
measure $P\left\{  \vec{r}(s)\right\}  \propto\exp\left[
-\mathrm{const} {\displaystyle\int} \left(
\partial^{2}\vec{r}/\partial s^{2}\right)  ^{2}ds\right]$,
subject to the constraint of non-extensibility, $\left\vert
\partial\vec{r}/\partial s\right\vert =1$. That simply means that
conformations are Boltzmann weighted by the bending energy
proportional to the squared curvature. In more general elastic
models \cite{RabinPanyukov}, the three generalized curvatures
describing the object could be treated as independent Gaussian
variables. However, this is true only for linear (with open ends)
filaments and attempts to generalize these methods to the case of
closed loop failed because the loop closure conditions introduce a
non-local coupling between the curvatures that makes the problem
practically intractable (except for the case of small fluctuations
of a planar ring in which this coupling can be explicitly taken
into account \cite{ring}). One natural way to generate the
conformations of a closed loop is to expand each component
$r^{i}\left(  \tau\right)  $ in a Fourier series:
\begin{equation}
r^{i}\left(  \tau\right)  =\sum\limits_{n=1}\left[
A_{n}^{i}\cos\left( \frac{2\pi n\tau}{T}\right)
+B_{n}^{i}\sin\left(  \frac{2\pi n\tau} {T}\right)  \right]  \ .
\label{eq:fourier}
\end{equation}
Here, $\tau$ is a parameter along the curve $\mathbf{r}(\tau)$
($0\leq\tau\leq T$) and the summation goes up to some cutoff
frequency $n_{\max}$ (if the series converges sufficiently
rapidly, the cutoff can be replaced by infinity).  Any
conformation of a sufficiently smooth closed loop can be fully
described by the set of Fourier coefficients $\left\{  A_{n}^{i},
B_{n}^{i}\right\}  $, giving rise to the concept of Fourier knots
introduced in \cite{Trautwein,Kauffman}. Notice that a Fourier
knot is a much more general concept than the more traditional
Lissajou knot, for which only one frequency is present for each
coordinate direction.

Using the above prescription one can generate an ensemble of loops of
different shapes by taking the Fourier coefficients from some random
distribution. However, even though all the loops have the same period $T$,
their contour lengths $L_{T}=\int_{0}^{T}d\tau^{\prime}\left\vert
d\mathbf{r/}d\tau^{\prime}\right\vert $ are different for each realization of
the Fourier coefficients. Even more importantly, since, in general,
$\left\vert \partial\vec{r}/\partial\tau\right\vert \neq1$, loops generated by
(\ref{eq:fourier}) do not obey the inextensibility condition and can not be
used to model worm-like polymers. In order to generate an ensemble of
different conformations of an inextensible loop of some well-defined contour
length, one has to transform to a representation in which trajectories are
parametrized by the arclength $s$
\begin{equation}
s\left(  \tau\right)  =\int_{0}^{\tau}d\tau^{\prime}\left\vert
d\mathbf{r/} d\tau^{\prime}\right\vert
,\quad\text{with\quad}\left\vert \partial\vec {r}/\partial
s\right\vert =1\label{eq:s}
\end{equation}
and then bring all the generated conformations to the same contour length by a
suitable affine transformation of all lengths and coordinates (this
transformation does not affect the topology since the latter is independent of
the parametrization). Thus, we can replace $L_{T}=s(T)$ by any standard length
$L$, provided that we rescale all lengths using the affine transformation:
$\mathbf{r}\rightarrow\mathbf{r}L/L_{T}$. Therefore, in three steps
(generation of random coefficients and Fourier summation (\ref{eq:fourier}),
reparametrization (\ref{eq:s}), and affine transformation $s\rightarrow
sL/L_{T}$), of which the latter two couple all Fourier harmonics together in a
complex way, we obtain a statistical ensemble of smoothly bent conformations
of an inextensible loop of length $L$. Let us now check if this ensemble is
representative of physical conformations of a worm-like loop which possesses
some characteristic bending and (possibly) twist rigidities.

The first step is to recognize that on length scales much larger
than some microscopic cutoff, the conformations of a worm-like
polymer are well represented by those of a BRW. For the latter,
the amplitudes of the Fourier components can be readily shown to
be of the form $\left\{  A_{n}^{i} ,B_{n}^{i}\right\}
_{BRW}=\lambda/n$ for $n\leq n_{\max}$ and zero otherwise, where
$\lambda$ is a random number (say, uniformly distributed between
$-1$ and $1$; the choice of another symmetric interval is
equivalent to rescaling the contour length, see below). As will be
demonstrated in the following, short scale behavior characteristic
of WLL can be obtained by replacing the abrupt cutoff $n_{\max}$
by Fourier coefficients that decrease smoothly with a
characteristic decay frequency $n_{0}$,
\begin{equation}
\left\{  A_{n}^{i},B_{n}^{i}\right\}  _{WLL}=\lambda
e^{-n/n_{0}}/n, \label{WLC coeff}
\end{equation}
(notice that the WLL and the BRW expressions for the coefficients coincide for
$n\ll n_{0}$).

\begin{figure*}
\centerline{\scalebox{0.9}{\includegraphics{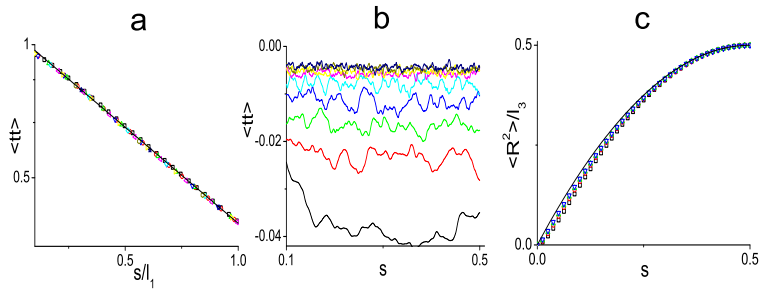}}}
\caption{(Color online) (a) Universal semi-log plot of the short
length scale behavior of the tangent-tangent correlation function
vs the scaled distance, $s/l_{1}$; the best fit
$exp(-(s/l_{1}-0.07))$ is given by the solid line. (b) Large scale
behavior of the correlation function, for different values of
$n_{0}$ (the height of the curves increases with $n_{0}$). (c)
Scaled (by $l_{3}$) mean-squared distance between points on the
loop. The BRW expression is plotted by the solid line. Different
values of $n_{0}$ are represented by black ($n_{0}=30$), red
($n_{0}=50$), green ($n_{0}=70$), blue ($n_{0}=100$), cyan
($n_{0}=150 $), magenta ($n_{0}=200$), yellow ($n_{0}=220$), dark
yellow ($n_{0}=250$) and navy ($n_{0}=270$). }
\label{fig:correlations}
\end{figure*}

The next step is to notice that the conformations of linear
worm-like polymer are governed by the persistence length, which
marks the cross-over between straight line conformations at small
length scales and random walk type behavior at large length scales
\cite{DeGennes}. We expect that on small scales, the conformations
of linear worm-like polymers and WLL are quite similar and that
for the latter this cross-over is represented in Fourier space in
terms of $n_{0}$. Of course, the analogy is meaningful only as
long as $n_{0}\gg1$, when there is a sufficiently broad range of
length scales between the persistence length and the contour
length of the knot. The characteristic property of the worm-like
chain model is exponential decay (as $\exp(-s/l_{1})$) of the
tangent-tangent correlation function, which defines the
persistence length $l_{1}$; similarly, for WLL, on length scales
sufficiently small compared to the entire loop, one expects that
$\left\langle \left\langle t(s)\cdot t(0)\right\rangle
\right\rangle \simeq\exp(-s/l_{1})$, where $\left\langle
\left\langle {}\right\rangle \right\rangle $ denotes averaging
both over the contour of a given loop and over the ensemble of
loops. This expectation is confirmed in Figure
\ref{fig:correlations}a, where the logarithm of the correlation
function is plotted against the dimensionless arclength $s/l_{1}$.
The choice $l_{1}=0.43/n_{0}$ allows us to superimpose data for
different values of $n_{0}$ in the range $30\leq n_{0}\leq270$
(the small shift of the exponent is the result of the finite
discretization of the contour length $s$).

At first sight, analogy with worm-like chain models of linear
polymers suggests that on length scales much larger than $l_{1}$
the conformations of the loops are those of BRW with step size
(Kuhn length) given by twice the persistence length, $2l_{1}$.
However, this analogy is open to question because the large scale
behavior of a loop is strongly affected by the loop closure
constraint and it is not clear, a'priori, whether the same
persistence length controls both the small and the large scale
behavior. We therefore decided to apply our method to generate a
representative ensemble of configurations of WLL and use it to
compute the tangent-tangent correlation function and the mean
squared distance between two well-separated points $s$ and
$s^{\prime}$ ($\left\vert s-s^{\prime}\right\vert \gg l_{1}$)
along the contour.

A simple estimate shows that on length scales much larger than the
persistence length, the tangent-tangent correlation function of a
WLL approaches a constant negative value $\left\langle
\left\langle t(s)\cdot t(0)\right\rangle \right\rangle
\simeq-2l_{2}/L$ where $l_{2}$ is, in general,different from
$l_{1}$ (the negative value of the correlation follows from the
fact that the tangent has to turn back on itself in order to come
back to its initial direction upon traversing the contour of the
loop). In Fig. \ref{fig:correlations}b this correlation function
is plotted in the interval $0.1<s<0.5$ (here and in the following
we take $L=1$) for different values of $n_{0}$. Upon averaging the
correlation functions over the oscillations, all the results for
different values of $n_{0}$ can be collapsed to a single
horizontal line by dividing them by $l_{2}=0.575/n_{0}$. We
therefore establish that unlike the case of linear worm-like
chains in which only a single persistence length exists, WLL have
at least two distinct persistence lengths, with
$l_{1}/l_{2}\simeq3/4$.

\begin{figure}
\centerline{\scalebox{0.6}{\includegraphics{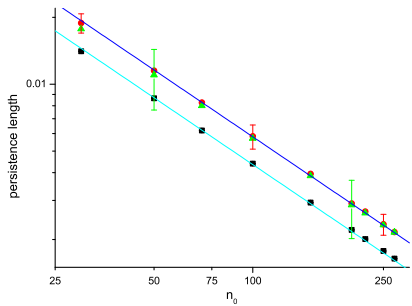}}}
\caption{(Color online).  Log-log plot of persistence lengths
$l_{1},l_{2},l_{3}$ (black squares, green triangles and red
circles respectively) vs $n_{0}$. The expressions $0.435/n_{0}$
and $0.57/n_{0}$ are shown as solid lines. The standard deviations
are represented by error bars.} \label{fig:persistence_lengths}
\end{figure}

We now proceed to compute the mean squared distance between two
points on the loop separated by a contour distance $s$,
$\left\langle \langle R^{2}\left( s\right)  \rangle\right\rangle
=(1/L)\left\langle \int_{0}^{L}\left\vert \mathbf{r}\left(
s^{\prime}+s\right)  -\mathbf{r}\left(  s^{\prime}\right)
\right\vert ^{2}ds^{\prime} \right\rangle $. Since we expect WLL
to behave like BRW on scales much larger than some persistence
length $l_{3}$, the probability distribution of $R^{2}(s)$ can be
easily written down assuming that both $s\gg l_{3}$ and $L-s\gg
l_{3}$. Apart from a normalization factor, this probability is
equal to the product of two Gaussian functions: $\exp\left[
-R^{2}/(2s\ell_{3})\right]  \exp\left[  -R^{2}/(2(L-s)\ell
_{3})\right]  $. Averaging with this distribution yields the well
known relation (see, e.g., \cite{RedBook}), $\left\langle
R^{2}\left(  s\right) \right\rangle _{\mathrm{BRW}}=2l_{3}s\left(
1-s/L\right)  .$ This result indicates that the plot of $\left.
\langle\langle R^{2}(s)\rangle \rangle\right/  l_{3}L$ against
$\sigma=s/L$ is universal, i.e., it is the unit height parabola
$4\sigma\left(  1-\sigma\right)$, independent of either
persistence length $l_{3}$ or total contour length $L$. Figure
\ref{fig:correlations}c shows that the data averaged over $1000$
different configurations collapse quite accurately on the expected
parabolic master curve.  Furthermore, by looking at our
computational data for $\left\langle \left\langle R^{2}\left(
s=L/2\right) \right\rangle \right\rangle $ averaged over all pairs
of diametrically opposite points on the loop ($s=L/2=0.5$) we were
able to relate the persistence length to the cutoff $n_{0}$.
Within the accuracy of our simulation $l_{3}$ coincides with
$l_{2}$ (see Fig. \ref{fig:persistence_lengths}).

The conclusion that local and global statistical properties of
worm-like loops are characterized by two different persistence
lengths, $l_{1}$ and $l_{2}(=l_{3}),$ respectively, is a new
result of this work. Although analytical theory of worm-like loops
does not exist at present, we can offer some tentative
considerations about the origin of the two persistence lengths,
based on the study of small fluctuations of elastic rings
\cite{ring}. For rings of zero linking number that possess both
bending and twist persistence lengths, it was shown that while on
small scales writhe fluctuations depend only on the bending
persistence length (on such scales the ring is essentially
straight and bending and twist decouple), both bending and twist
persistence lengths contribute to the writhe on larger scales, for
which the global geometry of the closed ring becomes important.
Extrapolating this result to the case of a strongly fluctuating
loop considered in this work, suggests that while $l_{1}$
represents the pure bending contribution to the persistence
length, $l_{2}$ contains both the bending and the twist
contributions and is therefore larger than $l_{1}$ \cite{Mezard}.

We now turn to the topological properties of WLL and examine the dependence of
the trivial knot probability $P_{0}$ on the cutoff frequency $n_{0}$. Since
$n_{0}$ determines the Kuhn length defined as either $2l_{1}$ or the $2l_{2}$,
we can bring our knotting probability data to a form comparable to that for
discrete polymer models, where trivial knot probability depends on the number
of segments $N$. In our case, we generate Fourier knots as completely smooth
curves, but we can define the number of effective segments as the ratio of the
contour length to the Kuhn length. For long random--walk-like polymers, this
definition coincides with the standard one accepted in polymer physics for the
Kuhn segments \cite{RedBook} but there remains an ambiguity associated with
the choice of $N_{1}=L/2l_{1}$ or $N_{2}=L/2l_{2}$. In order to resolve this
ambiguity we will measure the probability to obtain a trivial knot as a
function of the cutoff frequency $n_{0}.$

To address the topology of the loops computationally, we employ the knot
analysis routine due to R. Lua \cite{lua} which identifies knots by computing
the Alexander polynomial invariant $\Delta(t)$ at one value of argument,
$t=-1$, and Vassiliev invariants of degrees two, $v_{2}$, and three, $v_{3}$.
This set of invariants is widely considered as powerful enough for reliable
identification of the trivial knot for all lengths achievable in practical
computations. The details of this topological routine are described in
\cite{lua}.

\begin{figure}
\centerline{\scalebox{0.6}{\includegraphics{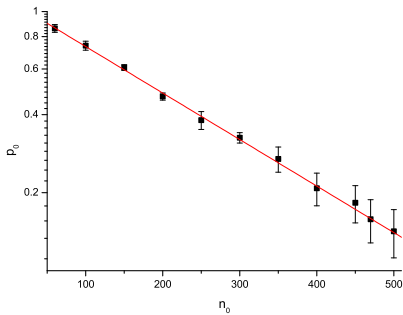}}}
\caption{(Color online).  Semi-log plot of the trivial knot
probability as a function of $n_{0}$.}
\label{fig:trivial_knot_probability}
\end{figure}

The data on the trivial knot probability are shown in Figure
\ref{fig:trivial_knot_probability}. As this figure indicates, the
trivial knot probability fits well to the exponential
\begin{equation}
P_{0}\left(  n_{0}\right)  \sim\exp\left[  -n_{0}/242\right]  \ .
\label{eq:exponential_fit}
\end{equation}
Notice that since this relation involves only the cutoff on the Fourier
series, formula (\ref{eq:exponential_fit}) can be re-interpreted in purely
mathematical form, not involving any references to polymers, or, for that
matter, to any physics. While there is no fundamental understanding of the
origin of this large cutoff ($n_{0}^{\ast}=242$) at present, our formulation
hints at the existence of a hitherto unexplored connection between Fourier
analysis and topology of space curves and will hopefully stimulate new work on
this fundamental problem.

In summary, we have shown that the choice of random Fourier
coefficients with amplitudes that decay with frequency as
$n^{-1}\exp(-n/n_{0})$, generates a statistical ensemble of
Fourier knots whose local properties coincide with those of
worm-like polymers with persistence length that scales as
$1/n_{0}$. We found that even though WLL behave on large scales as
BRW as expected, the effective step size of this BRW is larger
than that calculated from the local persistence length of the
loop. This new prediction can be directly tested by experiments on
double stranded DNA loops that can monitor both local (persistence
length) and global (e.g., radius of gyration) properties of the
polymers (see, e.g., \cite{AFM1,AFM2,SFM}). We also demonstrated
that similar to discrete models of polymers or random polygons,
our Fourier knots exhibit exponential decay of the unknotting
probability with the number of \emph{effectively} straight
segments, or, equivalently, with the maximal spatial frequency
included in Fourier expansion. The characteristic cross-over
determined by this exponential decay represents a large number,
which, although in the same ballpark as for other known models,
remains an unexplained puzzle.

\acknowledgements We acknowledge Ronald Lua's help in the use of his
computational knot analysis routine. The work of AG was supported in part by
the MRSEC Program of the National Science Foundation under Award Number
DMR-0212302. This research was supported in part by a grant from the US-Israel
Binational Science Foundation. YR would like to acknowledge the hospitality of
Institute for Mathematics and its Applications of the University of Minnesota.


\begin{thebibliography}{99}
\bibitem{Mezard} This effect is expected only for worm-like loops; for linear
worm-like chains it can be shown that twist degrees of freedom can be
integrated over and that the spatial conformations of the polymer depend on
the bending rigidity only: C. Bouchiat and M. Mezard, \emph{Eur. Phys. J.} E
\textbf{2}, 377 (2000).
\end{thebibliography}
\end{document}